# Observation of the *Memory Steps* in Graphene at Elevated Temperatures

*Sergey L. Rumyantsev[1,2*], Guanxiong Liu[3], Michael S. Shur[1] and Alexander A. Balandin[3]*

[1]Center for Integrated Electronics, Department of Electrical, Computer and Systems Engineering, Rensselaer Polytechnic Institute, Troy, NY 12180 USA

[2]Ioffe Institute, The Russian Academy of Sciences, St. Petersburg, 194021 Russia

[3]Nano-Device Laboratory, Department of Electrical Engineering and Materials Science and Engineering Program, Bourns College of Engineering, University of California – Riverside, Riverside, CA 92521

Corresponding authors (SLR): roumis2@rpi.edu

# ABSTRACT

We found that the current-voltage characteristics of the single-layer graphene field-effecttransistors exhibit an intriguing feature – an abrupt change of the current near zero gate bias at elevated temperatures $T > 500$ K. The strength of the effect – referred to as the *memory step* by analogy with the *memory dips* – known phenomenon in electron glasses – depends on the rate of the voltage sweep. The slower the sweep – the more pronounced is the step in the current. Despite differences in examined graphene transistor characteristics, the *memory step* always appears near zero gate bias. The effect is reproducible and preserved after device aging. A similar feature has been previously observed in electronic glasses albeit at cryogenic temperatures and with opposite dependence on the rate of the voltage sweep. The observed *memory step* can be related to the slow relaxation processes in graphene. This new characteristic of electron transport in graphene can be used for applications in high-temperature sensors and switches.

KEYWORDS: graphene, hysteresis, memory step, memory dip

Unique properties of graphene, such as extremely high carrier mobility[1,2] and intrinsic thermal conductivity[3,4], chemical inertness and mechanical stiffness[5] have attracted enormous attention. Many, although not all, unique features of this material have been understood. Practical applications of graphene in transparent electrically conductive electrodes and thermal management seem to be rather realistic. Far less clear is how to capitalize on graphene's properties in electronic and related applications. Here, we report on a new electronic effect in graphene, which we observed at elevated temperatures ($T > 500$ K). This effect can be a signature of some slow relaxation process or phase transitions in the graphene electronic system. In the vicinity of this transition, the sample is extremely sensitive to external perturbations, and therefore this effect might be very useful for electronic high temperature sensing applications. The possibility of practical applications and the hope to stimulate theoretical research in this field motivated this report of the phenomenological description of the observed intriguing phenomenon in graphene at elevated temperatures.

**Graphene Device Fabrication and Characterization**

Graphene samples were produced by mechanical exfoliation from bulk highly oriented pyrolytic graphite (HOPG). Graphene flakes, placed on standard Si/SiO$_2$ substrates, were identified using micro-Raman spectroscopy through the 2D band deconvolution and I(G)/I(2D) intensity comparison. We have reported details of our Raman measurements elsewhere[6,7]. The 8 nm Ti/80 nm Au (10 nm Cr/100 nm Au) metal layers were sequentially deposited on graphene by the electron-beam evaporation to produce the drain and source contacts[8,9]. The degenerately doped Si substrate acted as a back gate. The current–voltage characteristics (drain current versus gate voltage) were measured using a semiconductor parameter analyzer (Agilent 4156B). The characteristics were determined at temperatures from 300 K to 540 K for five single layer graphene (SLG) transistors. The measurement protocol was as follows: at the drain voltage $V_d$=100 mV, the gate voltage $V_g$=-40 V was applied and kept constant for 2 seconds. Then the gate voltage was swept from -40 V to 80 V and back to -40 V. In some cases, the single sweep from -40 V to 20 V was used.

Figure 1 shows three examples of the input current–voltage characteristics of graphene transistors at different temperatures. For all examined transistors, the charge neutrality point at room temperature ranged from 10V to 40 V. All the samples demonstrated a hysteresis for the direct and reverse gate voltage sweeps.

The inset in Figure 1 shows the effective mobility in graphene as a function of temperature. The effective mobility at $V_g$=-40 V as a function of temperature was calculated as

$$\mu_{eff} = \frac{L_g}{R_{Ch} C_g (V_{GS} - V_D) W}, \qquad (1)$$

where $V_{GS}$ is the intrinsic gate-to-source voltage, $V_D$ is the gate voltage corresponding to the minimum of current at charge neutrality point, $L_g$ is the transistor gate length, $C_g$ is the gate capacitance per unit area, $W$ is the gate width; $R_{Ch}$ is the channel resistance (see Ref. [9] for details of the calculation). With the temperature increase, the hysteresis became more pronounces, and the difference between the gate voltages corresponded to the current minimum increased.

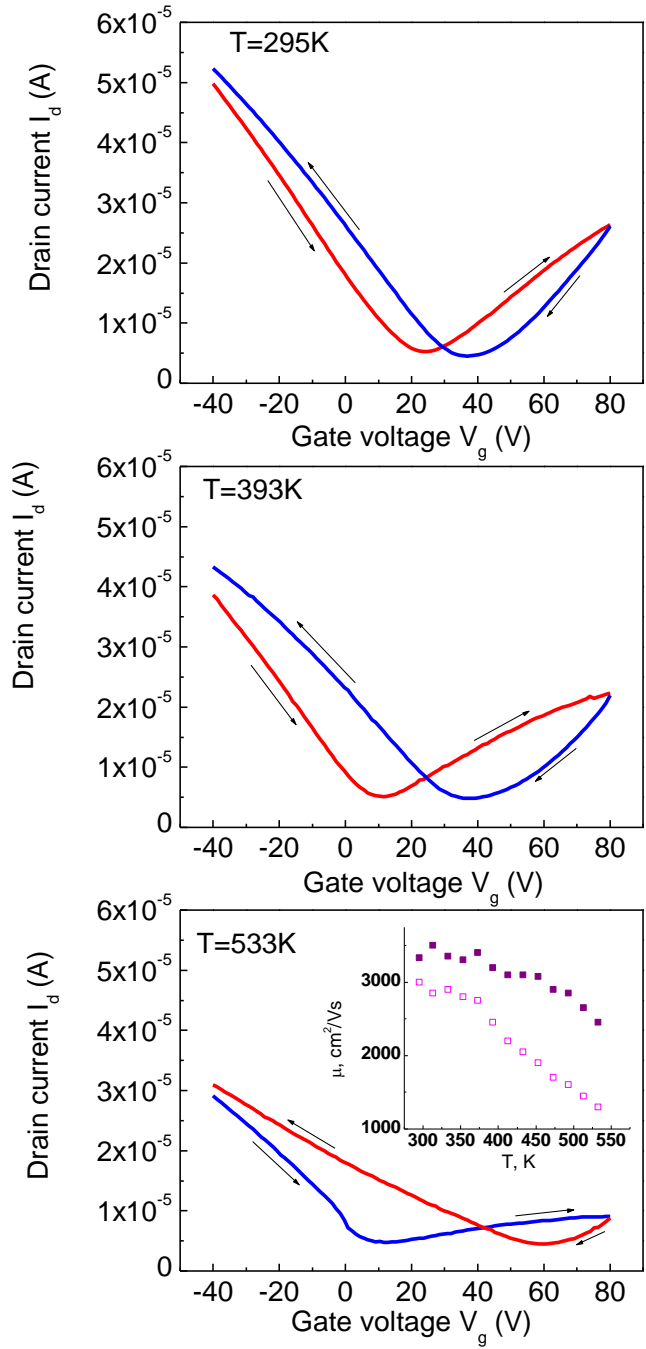

**Figure 1:** Current-voltage input characteristics of graphene back-gate transistor at different temperatures. The inset in the top panel shows a typical back-gated single-layer graphene transistor; the bottom panel inset is the temperature dependence of the effective mobility in graphene. Filled and open symbols show the mobility extracted at $V_g$=-40 V for the direct and reverse sweeps, respectively.

The hysteresis of current-voltage characteristics in graphene has been studied in several publications[10,15]. It is linked to slow carrier relaxation processes in graphene. These processes depend on temperature and can be affected by contamination from ambient air and the substrate. Such slow relaxations occur in many solid state systems due to different mechanisms, such as slow trapping and de-trapping of carriers by deep levels with small capture cross sections, slow change in the scattering cross-section, and the electron glass behavior[16-19].

As seen in Figure 1, in spite of the strong change in the current-voltage characteristics with temperature T, the current at the charge neutrality point only weakly depends on temperature. For the majority of the tested devices, the minimum conductivity remained within the range $(5.4\text{-}6.4)e^2/h$ $\Omega$.

**The Memory Steps in Graphene at Elevated Temperatures**

At temperatures above 500 K (or 220 $^o$C), we found a new feature in the current-voltage characteristic at $V_g \approx 0$. At the fast gate voltage sweep, it looked like a very small change of the slope of the drain-current versus the gate-voltage dependence. At the slow gate-voltage sweep, a clear step formed on the current-voltage characteristics (see Figure 2 that shows the current-voltage characteristics measured with $dV_g/dt=4.15$V/s and $dV_g/dt=0.43$V/s for one of the transistors.)

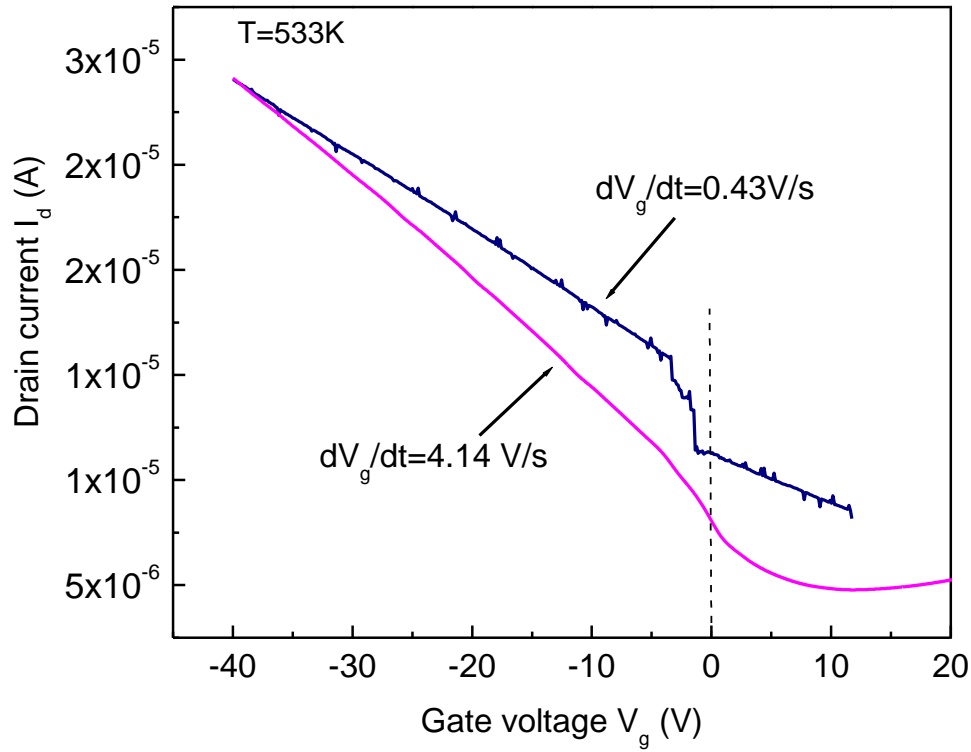

**Figure 2**: Current-voltage characteristics of graphene back-gate transistor at T=533 K measured at two different speeds of the voltage sweep $dV_g/dt$.

As one can see, the shape of the curves strongly depends on $dV_g/dt$. The slower is the scan, the more pronounced is the step on the slope. Here and below we refer to the feature as the "step" using an analogy with a similar feature in electronic glasses. Cooling the transistor and heating it again reproduced the I-V dependence including the step.

Figure 3 shows the evolution of the characteristics with temperature for $dV_g/dt$=0.43V/s.

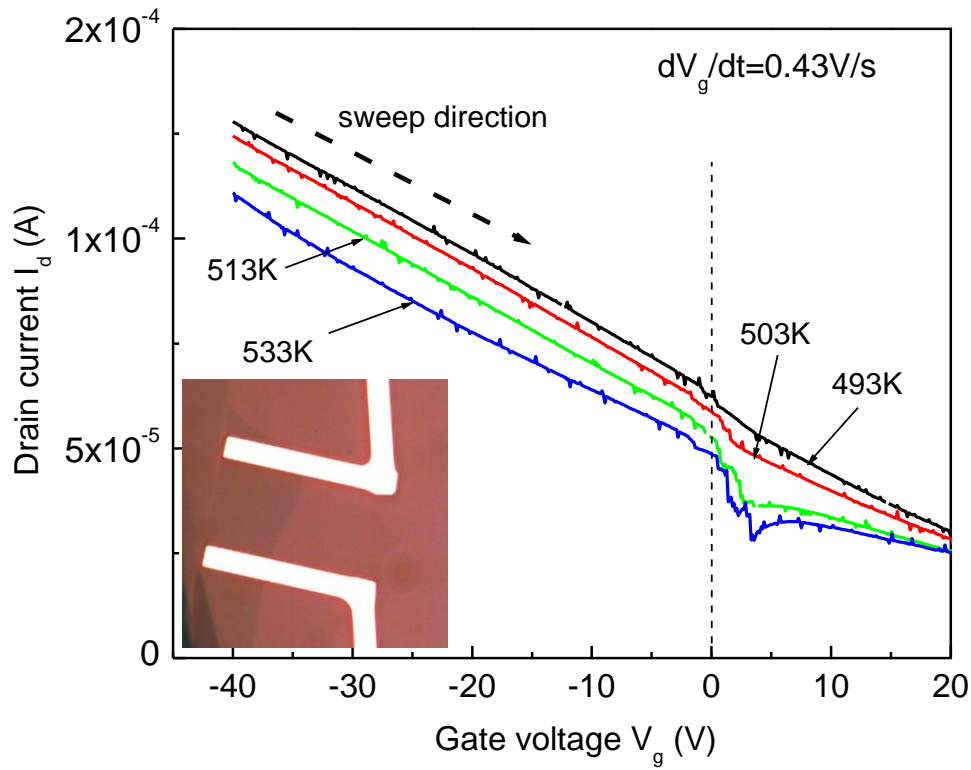

**Figure 3:** Current-voltage characteristics of graphene back-gate transistor at different temperatures. The inset shows the examined back-gated graphene device.

It is important to emphasize that all graphene samples demonstrated this kind of the step at the gate voltage close to but not exactly equal to zero despite having different shapes of the graphene flakes, conductivities, and the charge neutrality points. Figure 4 shows the current voltage characteristics for several samples. As seen, the voltage corresponding to the step varies slightly from sample to sample remaining close to $V_g=0$. For some samples, several steps were observed, which, however, did not reproduce well from the sweep to sweep.

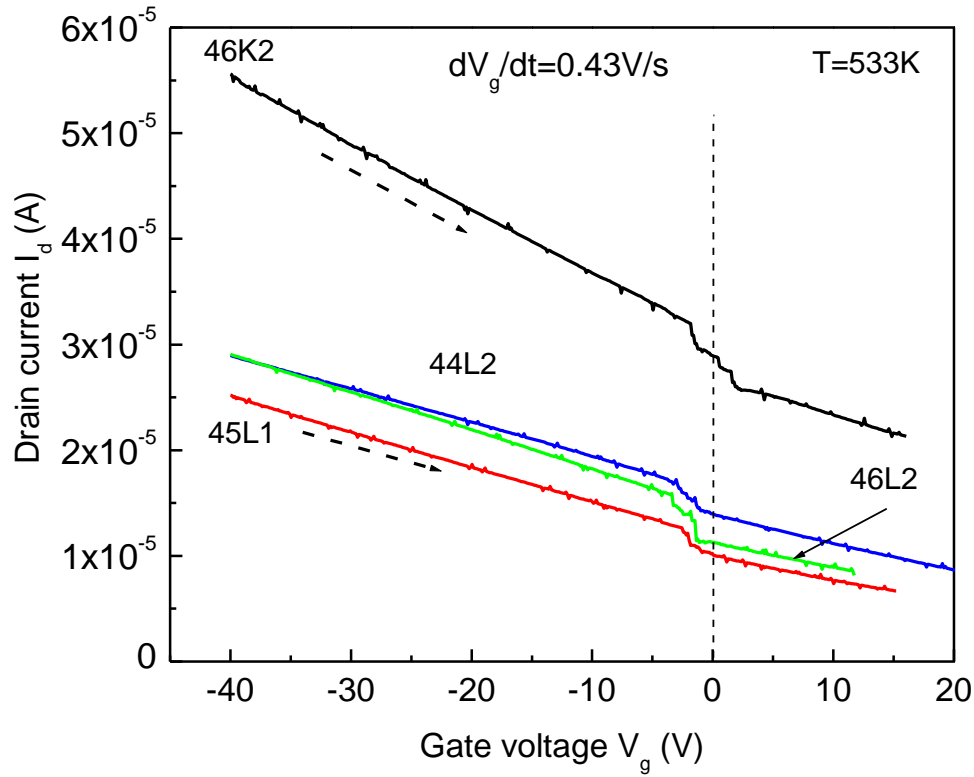

**Figure 4:** Memory steps at T=533 K (T=260 °C) observed in several different graphene transistors. Note that the step occurs near $V_g \approx 0$ V despite the fact that graphene transistors had different shape, charge neutrality point and electrical conductivity.

In order to verify how robust and reproducible this feature in I-V characteristic of graphene at elevated temperature, we kept the transistors at ambient environment for two weeks. As a result of aging, the conductivity dropped but the steps at T > 493-503 K (220 - 230 °C) were still clearly observed for all graphene devices.

To further test the reproducibility, we prepared another batch of back-gated graphene transistors, which had four in-plain contacts (see inset in Fig.5a) and which had distinctively different I-V characteristics. These transistors had much smaller hysteresis at room temperature and the Dirac voltage was within a few volts from $V_g$=0 V. The transistors have shown stable characteristics over sufficiently long time (about one month). The memory step was observed again although at slightly lower temperature and of slightly different shape. The striking feature was that the memory step was again at $V_g=0$ V (see Figure 5).

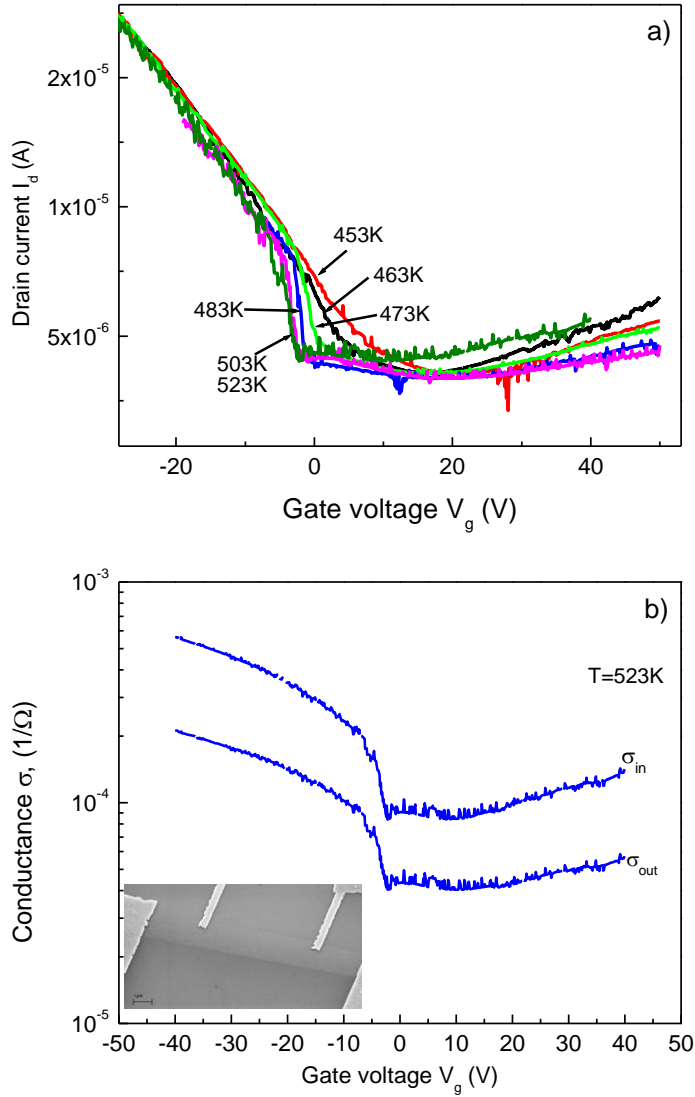

**Figure 5:** (a) Current-voltage characteristics measured in the four-probe configuration for the back-gate transistors at elevated temperatures. Despite the differences in the device performance and Dirac point position, the memory step was again observed near $V_g=0$ V. (b) The conductance measured between outer and inner contacts in the four-probe measurements. The inset shows the examined back-gated graphene device.

Figure 5b shows the conductance measured between outer, $\sigma_{out}$, and inner, $\sigma_{in}$, contacts in the four-probe configuration scheme. As seen $\sigma_{out}$, and $\sigma_{in}$, depend identically on the gate voltage indicating that the observed memory step does not relate to the contacts.

**Slow Conductance Relaxation and Memory Steps**

The exact origin of the observed memory step effect in graphene devices with different charge neutrality point and the reasons for its intriguing appearance near $V_g$=0 are not clear. At the same time, it is difficult not to notice a similarity between the observed feature in graphene and that studied in electron glasses[16-19]. The electron glasses demonstrated memory effects including the so-called memory dips, i.e., irregularity, referred to as the "dip", in the current versus gate-voltage dependences. Several mechanism responsible for the memory dip have been discussed (see [16 - 18] and references therein). However, the underlying physics of the memory dip in the electronic glass is still not yet well understood. In both systems, graphene and electron glass, this irregularity is found at $V_g$=0 and its amplitude is very sensitive to temperature[19]. However in contrast to our experiments, the memory dips in electron glasses are observed at low (cryogenic) temperatures and the amplitude of the dip increases with the increase of the gate voltage scan speed.

According to the theories that describe memory dips in electron glasses, many processes in solid-state systems involve slow conductance changes, which suggest non-equilibrium phenomena[16]. In degenerate Fermi systems, e.g. metals and heavily doped semiconductors, the fluctuations in the conductance $G$ (=$1/R$) reflect the temporal changes in the potential experienced by the charge carriers[20-21]. Such potential fluctuations may be structural, involving slow dynamics of atoms which, in turn, may be triggered by a modified state of local charge distribution. A slow release or trapping of carriers, likewise, manifest itself in the slow conductance fluctuations[16]. Either mechanism may lead to the conductance fluctuations that extend to very low frequencies and reveal themselves in the low-frequency spectrum dominated by $1/f$ and generation – recombination (G-R) noise ($f$ is the frequency). There are other mechanisms, which may lead to $G$ changing slowly with time. This may occur, for example, due to annealing of defects, diffusion of injected dopants, light exposure, irradiation, and other instances involving changes in the potential landscape, or the density of carriers in the conducting system. In such cases, the slow response observed in $G$ exhibits features that are characteristic of the electronic glasses[16].

We have previously observed the signature of slow trapping – emission processes in the low - frequency noise data from graphene transistors[9,22]. Specifically, from G-R peaks at very low $f$, we extracted time constants $\tau > 1$ second[9]. Other relaxation processes in graphene may take hours or even weeks. For example, we observed a slow recovery of electronic properties, e.g. mobility, of the electron-beam irradiated graphene when the samples were annealed or left for a long time in vacuum or ambient conditions[23-25]. In the present work, we found the memory steps at elevated temperatures ($T > 500$ K), which are high enough to trigger annealing processes, and slow relaxation resulting in electrical conductance change. Other graphene–specific slow relaxation processes can be related to the topological corrugations, which have been proved to exist in graphene and might be responsible for its stability[5]. The topological corrugations on the nanometer and micrometer scale can change slowly, particularly as the temperature changes from room temperature to $T > 500$ K.

The characteristic memory dips in electrical conductance in electronic glasses were also associated with the hopping type of electron conductivity[18]. This particular mechanism can also have its analog in graphene. It is known that in graphene, there exist simultaneously electron and hole puddles[26-27]. They are the result of the inevitable presence of disorder, which leads to emergence of the electron-rich and hole-rich regions. These puddles are likely the reason for the anomalous non-zero minimal conductivity at zero average carrier density[26]. The puddles are considered to be among factors limiting graphene mobility[27]. One can envision that electron and hole hopping among the puddles and slow evolution of the puddles themselves lead to the slow $G$ changes on the time-scale required for the observation of the memory steps.

**Conclusions**

We reported on a "memory step" occurring in the current-voltage characteristics of graphene field-effect transistors at elevated temperatures. The magnitude of the step depends on the rate of the voltage sweep. This new effect is observed in large variety of graphene samples. The effect is reproducible and preserved in the intentionally aged graphene devices. The observed step might be indicative of the reversible slow relaxation processes in graphene. The effect can be used for graphene applications in high-temperature sensors and switches.

*ACKNOWLEDGMENTS*

The work at UCR was supported, in part, by SRC – DARPA through the FCRP Functional Engineered Nano Architectonics (FENA) center. The work at RPI is supported by the NSF under the auspices of the I/UCRC "CONNECTION ONE." The authors also thank Professors Yuri Galperin, Valentin Kachorovskii, and Michael Levinshtein for helpful discussions.